\documentclass[pra,twocolumn]{revtex4-1}

\usepackage{graphicx}

\begin{document}

\title{Optoelectronic cooling of mechanical modes in a semiconductor nanomembrane}

\author{K.~Usami$^{1}$}
\author{A.~Naesby$^{1}$}
\author{T.~Bagci$^{1}$}
\author{B.~Melholt Nielsen$^{1}$}
\author{J.~Liu$^{2}$}
\author{S.~Stobbe$^{2}$}
\author{P.~Lodahl$^{2}$}
\author{E.~S.~Polzik$^{1}$} \email{polzik@nbi.dk}

\affiliation{$^{1}$QUANTOP - Danish National Research Foundation Center for Quantum Optics, Niels Bohr Institute, Blegdamsvej 17, 2100 Copenhagen, Denmark\\
$^{2}$DTU Fotonik, Department of Photonics Engineering, Technical University of Denmark, \O rsteds Plads 343, DK-2800 Kgs.\ Lyngby, Denmark}

\date{\today}

\begin{abstract}

Optical cavity cooling of mechanical resonators has recently become a research frontier~\cite{KV2008s,FK2009np,MG2009p}. The cooling has been realized with a metal-coated silicon microlever via photo-thermal force~\cite{MK2004n} and subsequently with dielectric objects via radiation pressure~\cite{GBPBLHSBAZ2006n,ACBPH2006n,SDNVK2006}. Here we report cavity cooling with a crystalline semiconductor membrane via a new mechanism, in which the cooling force arises from the interaction between the photo-induced electron-hole pairs and the mechanical modes through the deformation potential coupling. The optoelectronic mechanism is so efficient as to cool a mode down to 4~K from room temperature with just 50~$\mu$W of light and a cavity with a finesse of 10 consisting of a standard mirror and the sub-wavelength-thick semiconductor membrane itself. The laser-cooled narrow-band phonon bath realized with semiconductor mechanical resonators may open up a new avenue for photonics and spintronics devices.

\end{abstract}

\maketitle

The first experimental demonstration of the cavity back-action cooling utilized the photo-thermal (bolometric) force to cool a gold-coated silicon microlever~\cite{MK2004n}. The force results from the differential thermal expansion between the silicon lever and the gold film and the time lag is provided by the time needed to diffuse the thermal energy along the lever~\cite{MK2004n,MFOK2008b}. Another cooling mechanism based on the radiation pressure force with dielectric objects~\cite{GBPBLHSBAZ2006n,ACBPH2006n,SDNVK2006} has recently become the mainstream approach with a potentail to reach the oscillator ground state~\cite{KV2008s,FK2009np,MG2009p}. With this approach the vibrational mode temperature reduction by up to a factor of 4,000 has been demonstrated~\cite{GHVCGSA2009nphy}.  Cooling down to the ground state via radiation pressure requires a high-finesse cavity and its experimental realization has been hindered by the difficulty to achieve good mechanical qualities and low-loss optical properties on a micro- or nanometer scale simultaneously. A way to mitigate this dilemma by separating optical and mechanical functionalities in the so-called membrane-in-the-middle approach has been demonstrated~\cite{TZJMGH2008n,WRPK2009}.

Applying the cavity cooling method to semiconductors offers interesting prospects. In optoelectronics semiconductor devices with engineered electronic and photonic band-structures have enabled, for example, strongly enhanced light-matter interactions for quantum photonics applications~\cite{HBWGAGFHI2007n,LSJTSKFL2008}. Semiconductors are also important  in spintronics since they can be made ferromagnetic by doping with magnetic ions~\cite{Ohno1998s} and embrace localized long-lived electron and hole spins~\cite{KDHBSAF2004n,GBDOSKKSPW2008n,BGDWKSPW2009s}. Studying the interaction between the laser-cooled mechanical modes and these rich internal degrees of freedom of semiconductors may lead to a variety of new device conceptions in photonics and spintronics. A theoretical proposal has been put forward to take advantage of the strong deformation potential coupling in semiconductors for cooling mechanical resonators with an embedded quantum dot instead of using a cavity~\cite{WZI2004}. There is also the so-called optical refrigeration scheme~\cite{SE2007np}, which has been predicted to work efficiently for semiconductors and to reduce the entire lattice temperature down to 10~K, though no net cooling has yet been observed. 

Driven by these attractive features of semiconductors we have manufactured a $160$-nm-thick crystalline gallium arsenide (GaAs) membrane (see Methods) and investigated its potential for cavity optomechanics. We have found an extraordinarily large optomechanical coupling and photo-induced mechanical damping in a simple setting where a standard mirror and the membrane form a cavity with a finesse of only 10. Here we provide evidence of the first experimental realization of cavity cooling of mechanical modes using internal electronic degrees of freedom of semiconductors.

The experimental setup and the structure of the fabricated GaAs membrane are summarized in Fig.~1. A 1-mW diode laser (975~nm) is used to probe the membrane oscillations via a beam deflection method with a two-segment photodiode. By feeding the RF signal from the photodiode to a spectrum analyzer the mechanical resonant frequencies can be identified. The frequency of the fundamental mechanical mode (the (1,1)-mode) is 23.4~kHz (Fig.~2a). The other resonant frequencies observed conform within roughly 10\% to a simple model of the membrane being a taut rectangular drumhead and the mode shapes are identified by using a laser Doppler vibrometer with a similar membrane. A Ti:Sapphire laser providing the opto-mechanical coupling is mode-matched to the cavity. The ringdown of the mechanical modes is observed by inducing the oscillations with an intensity-modulated cavity field (see Methods). Figure~2b shows the results of the mechanical ringdown indicating that the damping rate is strongly enhanced by the cavity field at the wavelength of 810~nm and is linearly proportional to the cavity input power offset as is expected from the damping model (see Methods). Also shown is the frequency shift as a result of the cavity field. 

Note that the damping of the mechanical oscillations can only be observed when the cavity resonant frequency $\Omega_{\mathrm{C}}/2\pi$ is detuned to the red from the laser frequency $\Omega_{\mathrm{L}}/2\pi$, i.e., $\Omega_{\mathrm{C}}<\Omega_{\mathrm{L}}$, as opposed to the case of the cavity cooling via the radiation pressure~\cite{KV2008s,FK2009np,MG2009p}. Optomechanical instabilities (mechanical parametric oscillations)~\cite{ACBPH2006n,SDNVK2006} set in when $\Omega_{\mathrm{C}}>\Omega_{\mathrm{L}}$ (even for the cavity input power as low as 5~$\mu$W) or when the cavity input power exceeds 50~$\mu$W (regardless of the cavity detuning) preventing us from using these conditions for the ringdown measurements. The instabilities with such a small photon number imply a substantial optomechanical coupling.

To assess the mechanical properties of the membrane the intrinsic mechanical damping rate $\Gamma_{0}$ and the quality factor $Q_{0} \equiv \omega_{0}/\Gamma_{0}$ are inferred by extrapolating the ringdown data down to zero cavity field (where $\omega_{0}/2\pi \equiv \nu_{0}$ is the extrapolated bare resonant frequency). A markedly high quality factor $Q_{0}=2.3 \times 10^{6}$ is found for the (3,1)-mode at the frequency of 59.5~kHz (Fig.~2c and 2d). The product $Q_{0}\nu_{0}$ for this particular mode reaches $1.4\times10^{11}$~Hz at room temperature. For GaAs resonators such high $Q_{0}\nu_{0}$ products have been only observed to-date at a cryogenic temperature~\cite{OIOSY2009apex,CWVGPZWPA2010ieee}; although two orders of magnitude higher value at room temperature has been reported for stressed SiN nanomembranes~\cite{WRPK2009}. This observation strongly encourages ongoing efforts to use crystalline semiconductors for nano- and micro-electromechanics (NEMS and MEMS)~\cite{HSS1994jmm} and optomechanics~\cite{OIOSY2009apex,CWVGPZWPA2010ieee}. 

If the damping is not accompanied by any extra fluctuations, i.e., is the cold damping~\cite{FK2009np}, the enhanced damping also implies cooling. The effective temperature can then be defined as $T_{\mathrm{eff}}=T/(\Gamma_{\mathrm{eff}}/\Gamma_{0})$ (see Methods). The ratio of the damping rate at 50~$\mu$W (just below the instability-onset power) to the extrapolated intrinsic one thus gives the maximum cooling factor $\Gamma_{\mathrm{eff}}/\Gamma_{0}$ in our setting. From Fig.~2c the cooling factor for the (1,1)-mode is found to be about 11. For the (3,1)-mode, holding the highest $Q_{0}$, the cooling factor is about 75 as shown in Fig.~2d, meaning that the effective temperature of the mode is reduced down to as low as 4~K from room temperature.

To prove that the observed damping really means cooling frequency-domain measurements have been also performed. Figure~3a shows the power spectra of the mechanical resonances (the (1,1)-mode) for different cavity input powers. The slight discrepancy between the resonant frequencies of ringdown measurements in Fig.~2c and those in the spectrum in Fig.~3a is attributed to day-by-day temperature fluctuation (about $\pm$~3$^{\circ}$C). The area of the resonance, which is proportional to the effective temperature $T_{\mathrm{eff}}$~\cite{MK2004n,MFOK2008b}, clearly becomes smaller as the cavity input power increases. The original area with no cavity light divided by the reduced area gives the ratio $T/T_{\mathrm{eff}}$, i.e., the cooling factor. These ratios are displayed in Fig.~3b for different cavity inputs (purple circles). The width of the resonance, on the other hand, reflects the damping rate since it is equal to $\Gamma_{\mathrm{eff}}/2\pi$. The cooling factors obtained from the width measurements $\Gamma_{\mathrm{eff}}/\Gamma_{0}$ are shown in Fig.~3b (orange circles) along with the corresponding cooling factors obtained from the ringdown measurements (red diamonds). The three methods agree with each other within a factor of 3 (see Methods). Note that the similar discrepancy between area and width measurements has been reported~\cite{TZJMGH2008n}.

Further evidence of the cavity cooling is provided by the dependence of the cooling factor on the cavity detuning (Fig.~3c). In the cold damping scenario the effective damping rate $\Gamma_{\mathrm{eff}}$ is proportional to $\nabla F_{\mathrm{ph}}$ (see Methods). Here the slope of the cavity resonance corresponds to $\nabla F_{\mathrm{ph}}$ since $F_{\mathrm{ph}}$ is presumably proportional to the intra-cavity photon number. The red solid line in Fig.~3c indicates the slope of the cavity resonance, i.e., $\nabla F_{\mathrm{ph}}$. Although the data from the ringdown measurements differ in detail from the expected behavior, it is evident that the enhancement of damping reflects $\nabla F_{\mathrm{ph}}$.

We attribute the observed very efficient cooling to the effect of above-band-gap photons. Photo-induced electrons in the conduction band (holes in the valence band) leads to an electronic stress in the lattice through the deformation potential coupling~\cite{TGMT1986b,MTFBW2005b}, because the conduction band electrons bind the lattice weaker than the valence band electrons. Growth of the cavity field due to the membrane motion leads to the photo-induced force on the membrane in the direction of motion, which is
opposite to the case of radiation pressure. This explains why the cavity
detuning corresponding to cooling in the present case is opposite to that
for radiation pressure cooling. The electronic stress $\sigma_{\mathrm{el}}=-B\frac{dE_{\mathrm{g}}}{dp}$ per single electron-hole pair generated by a photon with the energy $E=$~1.53~eV (810~nm) can be 100 times larger than the corresponding thermal stress $\sigma_{\mathrm{th}}=-B\frac{3\beta}{C}(E-E_{\mathrm{g}})$, where the band gap of GaAs at room temperature is $E_{\mathrm{g}}$~=~1.43~eV (865~nm)~\cite{MTFBW2005b}. Here, $B$ is the bulk modulus, $-B\frac{dE_{\mathrm{g}}}{dp}$ is the hydrostatic deformation potential, and $\beta$ and $C$ denote the thermal expansion coefficient and the heat capacity, respectively~\cite{MTFBW2005b}. 
 There is also the piezoelectric coupling in GaAs, but compared to the deformation potential coupling its contribution to the cooling is negligible~\cite{RRS1990b}.

The presented optoelectronic cooling mechanism is supported by the observed dependence of the cooling factor on the photon energy (wavelength) of the cavity field in the range from $E=$~1.53~eV (810~nm) down to 1.40~eV (884~nm) across the band gap $E_{\mathrm{g}}$. The results are displayed in Fig.~4. Firstly, the cooling factor is essentially constant from $E\gg E_{\mathrm{g}}$ to $E\sim E_{\mathrm{g}}$. This observation rules out the possibility of the photo-thermal force being responsible for the cooling since it explicitly depends on $E-E_{\mathrm{g}}$. The fact that the expected stress ratio $\sigma_{\mathrm{th}}/\sigma_{\mathrm{el}}$ is, as mentioned above, around 0.01 for $E=$~1.53~eV (810~nm) and it drops further as $E$ approaches $E_{\mathrm{g}}$ also suggests that the contribution of $\sigma_{\mathrm{th}}$ to the cooling is marginal. Secondly, the cooling gradually reduces when $E \sim E_{\mathrm{g}}$ and ceases to effect when $E < E_{\mathrm{g}}-k_{\mathrm{B}}T$ ($k_{\mathrm{B}}$ is the Boltzmann constant and $k_{\mathrm{B}}T\sim$~26~meV at room temperature), supporting the assertion that the carrier excitation is responsible for the cooling. Last, but not least, we find a striking similarity between the trajectory of the cooling factor in Fig.~4 and the photoconductivity response of a similar GaAs membrane with an embedded field-effect transistor ~\cite{HELFB1995ssc}. The slight decline of the cooling factor starting around 1.49~eV could be due to the interaction of the photo-excited carriers with the optical phonons as seen in the photoconductivity measurement~\cite{HELFB1995ssc,Nahory1969pr}.

The observed dependencies of $\omega_{\mathrm{eff}}/2\pi$ and $\Gamma_{\mathrm{eff}}$ on optical power (Fig.2) can be used to infer $\tau$, the delay time between the photo-induced force and displacement~\cite{MFOK2008b}, which is specific for every cooling mechanism~\cite{KV2008s,FK2009np,MG2009p}. Using equations $\omega_{\mathrm{eff}}/2\pi$ and $\Gamma_{\mathrm{eff}}$ given in the Methods section we determine $\nabla F_{\mathrm{ph}}/(m\omega_{0}^{2})\sim 0.01$ and $\tau \sim$~10~ns. This fairly long delay time $\tau$ can be backed by the long ($\tau_{e}\ge 10$~ns) radiative lifetime $\tau_{e}$ of free excitons in GaAs~\cite{tvMF1987b}, though more studies are required to decisively determine the delay time $\tau$. An outstanding issue is the onset of the instability just above 50~$\mu$W which is unexpected from the simple model that we use. There are at least three distinctive time scales involved in the cooling mechanism; the mechanical oscillation time $1/\omega_{0}$, the cavity decay time $1/\kappa$, and the delay time $\tau$ introduced by the internal degrees of freedom. This clearly calls for additional theoretical and experimental work to investigate these effects in the context of cavity optomechanics.  

The demonstrated optomechanical coupling is very efficient. Indeed significant changes in the membrane dynamics are observed already with a few microwatts of the input power corresponding to just 10,000 intracavity photons for our 3-cm cavity.  For a 300-$\mu$m cavity this number is down to 100 photons and even this number can be probably further reduced by a suitable design. Using a higher finesse cavity with the membrane in the middle, the optical power can also be considerably reduced.

\section*{Methods}

\textbf{GaAs membrane fabrication.}
The GaAs membrane is fabricated from a GaAs/AlGaAs heterostructure wafer comprised of a (100)-oriented GaAs substrate (350~$\mu$m), an Al$_{0.85}$Ga$_{0.15}$As etch stop layer (1~$\mu$m), and a GaAs capping layer (160~nm) as shown in Fig.~1b. The backside of the wafer is patterned by photolithography and a selective wet etching process using a citric-acid-based solution is employed to remove the GaAs substrate in the area selected by the photolithographic pattern. Finally, another selective wet etching process using hydrofluoric acid is used to remove the sacrificial layer, thus resulting in a 160-nm-thick suspended GaAs membrane as shown in Figs.~1b and c.

\textbf{Ringdown measurements.}
A Ti:sapphire laser operating in the range from 810~nm to 884~nm is mode-matched to the cavity inside a vacuum chamber and is used for both cooling and inducing mechanical oscillations by modulating the intensity at frequency $\omega/2\pi$ around a specific offset level. The cavity length ($\sim$~29~mm) can be stabilized to a variable position on the cavity resonant slope. The RF signal from the two-segment photodiode and a reference signal at frequency $\omega/2\pi$ from the signal generator are fed into a 2-phase lock-in amplifier as shown in Fig.~1a. The produced mixed-down signals $\mathrm{x_{cos}}$ and $\mathrm{x_{sin}}$ correspond to the $\cos(\omega t)$ and $\sin(\omega t)$ components of the membrane displacement within the bandwidth of $d\omega/2\pi=$~500~Hz centred around $\omega/2\pi$, and their squared sum corresponds to $|x_{\omega}|^{2}d\omega$. For the ringdown measurements a digital oscilloscope is used to track the signal $|x_{\omega}|^{2}d\omega$ triggered by a TTL signal, by which the modulation of the cavity field is also turned off. The ringdown time of the mechanical oscillations is measured starting from the moment when the modulation is turned off while keeping the offset power level of the cavity input for cooling. 

\textbf{Cold damping.}
In the cold damping scenario~\cite{FK2009np} the equation of motion for the membrane displacement in the Fourier space is modified into $-\omega^{2} x_{\omega} + i \omega \Gamma_{\mathrm{eff}} x_{\omega} + \omega_{\mathrm{eff}}^{2} x_{\omega} = F_{\mathrm{th}}/m$ due to the photo-induced force $F_{\mathrm{ph}}$, where $F_{\mathrm{th}}$ is the thermal Langevin force and $m$ is the motional mass of the membrane~\cite{MK2004n,MFOK2008b}. Here the modified resonant frequency $\omega_{\mathrm{eff}}/2\pi$ and the effective damping rate $\Gamma_{\mathrm{eff}}$ can be expressed as $\omega_{\mathrm{eff}}^{2}=\omega_{0}^{2}(1-\frac{1}{1+\omega_{0}^{2}\tau^{2}}\frac{\nabla F_{\mathrm{ph}}}{m \omega_{0}^{2}}))$ and $\Gamma_{\mathrm{eff}}=\Gamma_{0}(1+Q_{0}\frac{\omega_{0} \tau}{1+\omega_{0}^{2}\tau^{2}}\frac{\nabla F_{\mathrm{ph}}}{m \omega_{0}^{2}})$, respectively. $\nabla F_{\mathrm{ph}}$ is the spatial derivative of $F_{\mathrm{ph}}$ introduced by the cavity and $\tau$ is the relevant time lag in the physical mechanism responsible for these modifications~\cite{MK2004n,MFOK2008b}. The effective temperature can then be defined as $T_{\mathrm{eff}}=T/(\Gamma_{\mathrm{eff}}/\Gamma_{0})$~\cite{MFOK2008b}.

\textbf{Time-domain and frequency-domain measurements.}
Since the mechanical oscillations arise merely out of the thermal Langevin force $F_{\mathrm{th}}$ for the frequency-domain measurements the resultant signal-to-noise ratio is generally worse than the ringdown measurements, for which the intensity-modulated cavity field can be used to amplify the oscillations. Note also that sub-Hertz mechanical resonances are measured with a spectrum analyzer with the resolution bandwidth of 1~Hz, which may contribute to the uncertainty of the determination of the area and certainly affects the width and height of the resonances. Because of these deficiencies of the frequency-domain measurements, the majority of our measurements are done in the time-domain method, i.e., the ringdown.

\section*{Figure Legends}
\textbf{Figure~1| Experimental setup and structure of the fabricated GaAs membrane.} \textbf{a}, Sketch of the experimental setup. A standard dielectric concave mirror with the reflectivity of 85\% and a fabricated 160-nm-thick GaAs membrane (see Methods) with the reflectivity of 75\% form a hemispherical cavity with the measured finesse of 10 in the range of the laser wavelength from 810~nm to 884~nm. The cavity is placed in a vacuum chamber kept at 10$^{-5}$~Pa. The cavity length ($\sim$~29~mm) can be varied with a piezoelectric transducer (PZT) attached to the end mirror. A Ti:sapphire laser is used as a cavity input for both cooling and inducing mechanical oscillations (the beam spot size at the membrane is about 80~$\mu$m in radius) by modulating the intensity with a signal generator and an acousto-optic modulator (AOM). A diode laser (975~nm) is used to probe the membrane oscillations via a beam deflection method with a two-segment photodiode. For ringdown measurements a 2-phase lock-in amplifier is used to produce a signal $|x_{\omega}|^{2}d\omega$ and a digital oscilloscope is used to track the signal (see Methods). \textbf{b}, Cross section of the fabricated 160-nm-thick suspended GaAs membrane (see Methods). \textbf{c}, Dimensions of the membrane. 

\textbf{Figure~2| Ringdown results.} \textbf{a}, \textbf{b}, Images of mode shapes for the fundamental mechanical mode (the (1,1)-mode) and the (3,1)-mode, respectively. \textbf{c}, \textbf{d}, Ringdown results with the cavity field at the wavelength of 810~nm for the (1,1)-mode and the (3,1)-mode, respectively. In \textbf{b} and \textbf{d} red and blue diamonds respectively represent the measured damping rates and the frequencies used to induce the mechanical oscillations. Each point is the average of 5 identical measurements and the error bars correspond to one standard deviation. Lines are produced by the least-square linear fits.

\textbf{Figure~3| Evidence of cooling.} \textbf{a}, Power spectra of the mechanical resonances (the (1,1)-mode) for different cavity input powers (the cavity field at the wavelength of 810~nm). \textbf{b}, Cooling factors obtained from the areas (purple circles) and the widths (orange circles) of the resonances. The cooling factors obtained from the ringdown measurements (shown in Fig.~2b) are also displayed with the error bars (red diamonds). Each line is produced by the least-square linear fit. For fitting the cooling factors obtained from the widths (orange circles) the first 4 points are not used since they are limited by the resolution bandwidth of the spectrum analyzer. \textbf{c}, Dependence of the cooling factor on the cavity detuning for the cavity input power of 30~$\mu$W. The cavity detuning is normalized to the cavity bandwidth $\kappa$. Red solid line indicates $\nabla F_{\mathrm{ph}}$ (see Methods). In \textbf{b} and \textbf{c} black dotted horizontal lines represent a cooling factor equal to one, i.e., no-cooling.

\textbf{Figure~4| Dependence of the cooling factor on the photon energy of the cavity field.} Each point corresponds to the cooling factor for the (1,1)-mode with a given intracavity photon number (equal to the number of 1.53-eV (810-nm) photons inside the cavity for the input power of 50~$\mu$W). Black dotted vertical line indicates the GaAs band gap energy $E_{\mathrm{g}}$~=~1.43~eV (865~nm), black dotted horizontal line represents a cooling factor equal to one, i.e., no-cooling. Conceptual illustrations of a electron-hole pair generation by a photon of the energy of $E > E_{\mathrm{g}}$ (left) and $E < E_{\mathrm{g}}$ (right) are shown on the bottom with the band structure. 

\section*{Acknowledgements}
We thank J.~Appel, A.~Grodecka-Grad, J.~H.~M\"{u}ller and D.~J.~Wilson for discussions, S.~Schmid for his technical support. This work was supported by the Japan Science and Technology Agency (JST), the Japan Society for the Promotion of Science (JSPS), the EU Project Q-ESSENCE, and the Danish Council for Independent Research (Technology and Production Science and Natural Science).

\section*{Author Contributions}
K.U., B.M.N. and E.P. designed the experiment. K.U., A.N. and T.B. worked on data collection and analysis. J.L. and S.S. fabricated the GaAs membranes. P.L. and E.P. planned and supervised the study. K.U., S.S. and E.P. wrote the manuscript. All authors discussed the results and commented on the manuscript.

\begin{figure}
\includegraphics[width=0.90\linewidth]{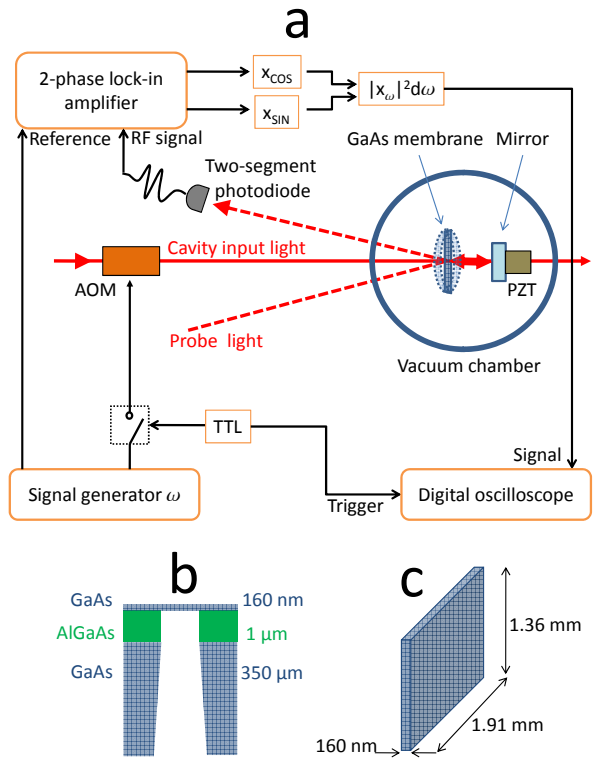}
\caption{\textbf{Experimental setup and structure of the fabricated GaAs membrane.}}
\label{fig1}
\end{figure}

\begin{figure}
\includegraphics[width=0.90\linewidth]{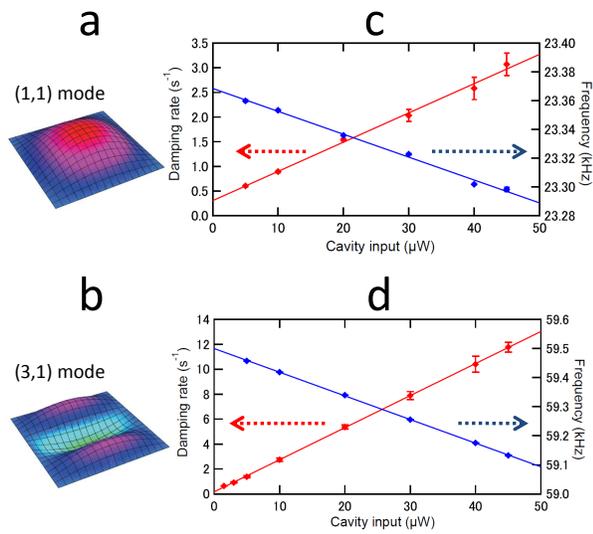}
\caption{\textbf{Ringdown results.}}
\label{fig2}
\end{figure}

\begin{figure}
\includegraphics[width=0.80\linewidth]{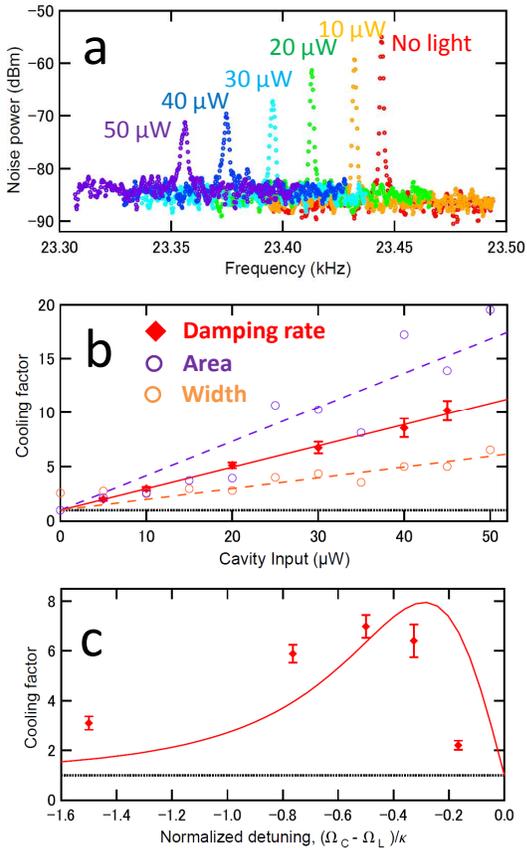}
\caption{\textbf{Evidence of cooling.}}
\label{fig3}
\end{figure}

\begin{figure}
\includegraphics[width=0.90\linewidth]{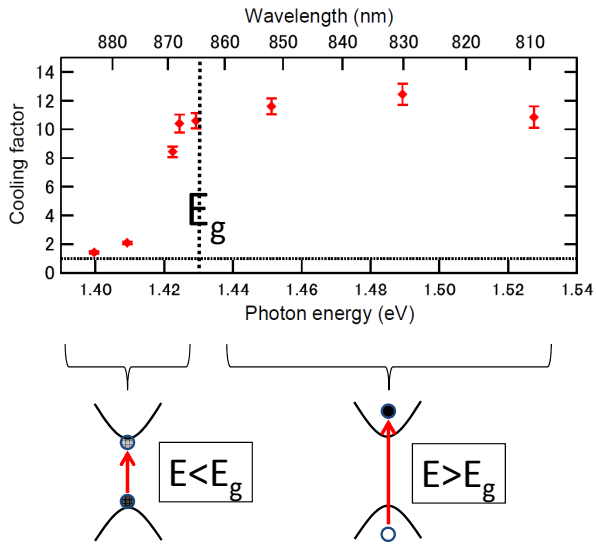}
\caption{\textbf{Dependence of the cooling factor on the photon energy of the cavity field.}}
\label{fig4}
\end{figure}

\end{document}